\documentstyle[12pt,epsfig]{article}
\begin{document}
\begin{titlepage}
\vskip -3em 
\flushright{BUHEP-96-48}
\vskip 5em
\begin{center}
{\Large\bf $B \to X_s \mu^+ \mu^-$ in Technicolor with Scalars}
\vskip 3em
{\large Yumian Su\footnote{email: {\tt yumian@buphy.bu.edu}}}
\vskip 2em
{\normalsize \it Dept. of Physics, Boston University,\\590 Commonwealth Ave.,
            Boston, MA 02215}
\vskip 3em 
{\large \today}
\vskip 6em

\begin{abstract}

    We compute the flavor changing neutral current process 
$B \to X_s \mu^+ \mu^-$ in technicolor models with scalars.
We find that the branching ratio can be enhanced relative
to that of the Standard Model by as much as $60\%$. The full
parameter space of the model is consistent with the present 
CLEO and CDF exclusive limits. However, the viability of the
model could soon be tested since the decay signals are 
expected to be observed in the next few years with the 
upgraded CLEO detector and in the CDF Run II.  

\end{abstract}
 
\end{center}  

\end{titlepage}

%%%%%%%%%%%%%%%%%%%%%%%%%%%%%%%%%%%%%%%%%%%%%%%%%%%%%%%%%%%%%%%%%%%%%%%%%%%

\section[short title]{Introduction} 

\bigskip  

Technicolor with scalars is a very simple and calculable kind of 
technicolor model, in which a scalar doublet with a positive mass
squared is introduced to couple the technifermions with the ordinary 
fermions.  When the technicolor interaction becomes strong and
technifermions condense, this scalar develops a vacuum expectation value
(VEV), which breaks the electroweak symmetry together with the 
technipions.  This scalar VEV is also responsible for giving masses 
to ordinary fermions.  It is very interesting that this model can be
treated as a kind of low energy effective theory of strongly-coupled ETC
(SETC) models~\cite{Sekhar:Cohen},  when some degree of fine tuning is 
allowed. This fine-tuning is necessary in any workable SETC model to
maintain a sufficient hierarchy between the ETC and technicolor 
scales~\cite{Lane}.  Some phenomenological issues have been explored 
previously in the literature and the model has been proved to be able
to stand the experimental tests so far~\cite{EHS}-\cite{CC:EHS:YM}.

  In this article, we consider the process $B \to X_s \mu^+\mu^-$ 
in the hope of testing the model and constraining its 
parameter space further.  The present exclusive limit on this channel 
from CDF,
$\mbox{BR}(B^0 \to {K^*}^0 \mu^+ \mu^-)_{\mbox{{\scriptsize CDF}}}<2.5
 \times10^{-5}$~\cite{CDF},
is within an order of magnitude of the Standard Model (SM) prediction,
$\mbox{BR}(B \to X_s \mu^+ \mu^-)_{\mbox{{\scriptsize SM}}}=(5.7\pm1.2)  
\times10^{-6}$~\cite{Ali}. The CLEO exclusive limit on the branching
ratio ( $3.1\times 10^{-5}$ ) is less stringent~\cite{CLEO}.

  In the SM, this process can only occur at loop level, and the error in
the evaluation of $\Gamma(B \to X_s \mu^+ \mu^-)$ can not be reduced to
less than $10 - 20\%$ due to the uncertainties in quark masses and 
the interference effects from excited charmonium states~\cite{Ligeti}. 
Still, given the large mass of top quark, one may expect this decay
to be sensitive to new physics contributions, if these contributions 
significantly overwhelm the QCD uncertainties. Therefore, the measurement of 
$B \to X_s \mu^+\mu^-$ can provide a probe of the validity of critical 
ingredients of the Standard Model and, possibly, of the existence of 
new physics beyond. If the branching ratio does not lie significantly below 
the SM prediction, positive signals are expected to be observed with
the upgraded CLEO detector and in the CDF Run II\footnote{CDF Run II is
expected to observe the decay of $B^0 \to {K^*}^0 \mu^+ \mu^-$ even if
its branching ratio is as low as $3.4 \times10^{-7}$~\cite{CDF:RunII}.}.

 In section 2, we present the technicolor with scalars model.  We then compute
 the $B \to X_s\mu^+\mu^-$ branching ratio and discuss the results in section 3.
Finally, in section 4 we give our conclusions. 

%%%%%%%%%%%%%%%%%%%%%%%%%%%%%%%%%%%%%%%%%%%%%%%%%%%%%%%%%%%%%%%%%%%%%%%%%%%%%%%%
\section[short title]{Technicolor with Scalars}

\begin{figure}
\vskip -3.5em
\centerline{\epsfig{file=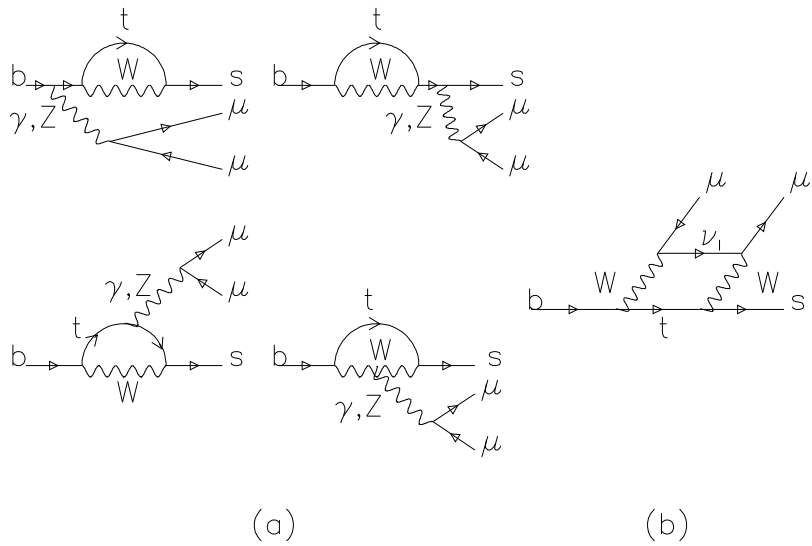,height=14cm,width=14cm}}
\vskip -2.5em
\caption{\label{smdiag} Penguin (a) and box (b) diagrams for 
         $b \to s \mu^+ \mu^-$ in the Standard Model. }
\end{figure}

\begin{figure}
\vskip -3.5em
\centerline{\epsfig{file=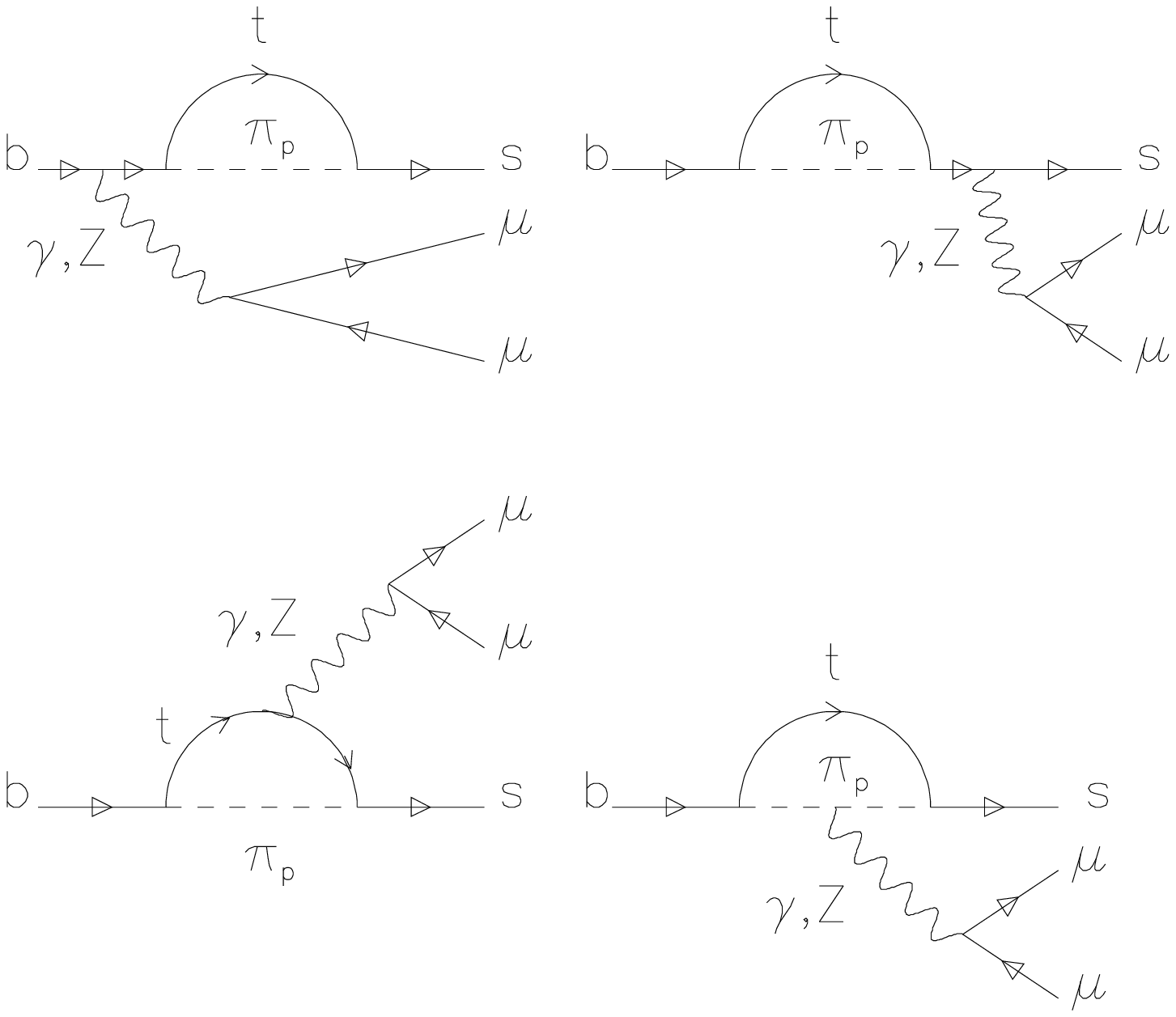,height=12cm,width=12cm}}
\vskip -2.5em
\caption{\label{tcsdiag}
   Additional Feynman diagrams contributing to $b \to s \mu^+
   \mu^-$ in technicolor with scalars. $\pi_p$ is the charged physical
  scalar in this model.}
\end{figure}

In the Standard Model, $B \to X_s \mu^+ \mu^-$ is dominated by 
one-loop contributions involving the exchange of a virtual $W$ and 
top quark (See figure \ref{smdiag}).  In the technicolor with scalars
model, there exists at least one physical charged scalar, which can be 
exchanged in the loop\footnote{Here, we ignore the 
   box diagram which can be obtained 
   from figure 1(b) by changing $W$ into $\pi_p$, because the $\pi_p$ 
   coupling to the muon is small (proportional to the muon mass).}, 
together with a top quark (figure \ref{tcsdiag}).
Therefore, we need to know the mass of this 
scalar and its interaction with quarks. 
%%%%%%%%%%%%%%%%%%%%%%%%%%%%%%%%%%%%%%%%%%%%%%%%%%%%%%%%%%%%%%%%
%%%%%%%%%%%%%%%%%%%%%%%%%%%%%%%%%%%%%%%%%%%%%%%%%%%%%%%%%%%%%%%%%%%%%%%%%
%%%%%%%%%%%%%%%%%%%%%%%%%%%%%%%%%%%%%%%%%%%%%%%%%%%%%%%%%%%%%%%%
%%%%%%%%%%%%%%%%%%%%%%%%%%%%%%%%%%%%%%%%%%%%%%%%%%%%%%%%%%%%%%%%%%%%%%%%%

We will now give a brief summary of the model, focusing on the fact 
that we need to determine these quantities needed  in our computation.
For more details of the model, we refer the reader to \cite{EHS} and
\cite{CC:HG}. 

The gauge group in the model is 
$SU(N)_{TC}\times SU(3)_C \times SU(2)_L \times U(1)_Y$, 
with ordinary particle content the same as in the 
Standard Model. These ordinary particles are singlets under the SU(N) 
technicolor group.  Let us now consider the simplest case where there
is a single weak doublet of technifermions
$$\upsilon_L=\left(\begin{array}{c} p\\m
                 \end{array}\right)_L ,\ \  p_R,\  m_R,$$
where their hypercharges
$Y(\upsilon_L)=0,Y(p_R)=\frac{1}{2},Y(m_R)=-\frac{1}{2}$ 
are chosen to cancel gauge anomalies.  In addition to the above 
particle spectrum, there exists a scalar doublet, $\phi$, 
\begin{displaymath}
\phi=\left(\begin{array}{c} \phi^+\\\phi^0
                  \end{array}\right),
\end{displaymath}
which has hypercharge 1/2.  This scalar couples to the technifermions, as well 
as ordinary fermions.  After the condensation of technifermions, 
because of the common scalar $\phi$, ordinary fermions obtain masses.

  The isotriplet scalar bound state of $p$ and $m$ (technipions), 
and the isotriplet components of $\phi$ will mix. One linear combination
becomes the longitudinal component of $W$ and $Z$, while the orthogonal 
combination remains in the low energy theory as an isotriplet of
physical scalars, $\pi_p$, whose coupling to quarks is~\cite{CC:HG}
\begin{equation}
i\left(\frac{f}{v}\right)\left[\bar{D}_LV^{\dag}\pi_p^-h_UU_R
+ \bar{U}_L\pi_p^+Vh_DD_R +h.c.\right].
\end{equation}
Here V is the CKM matrix, f is the technipion decay constant,
and $v$ is the electroweak scale ($ \approx 250$GeV); $U$, and $D$ are 
column vectors in flavor space; $h_U$ and $h_D$ are Yukawa coupling 
matrices.  The above looks like the interaction of Higgs doublet with
quarks in a type-I two Higgs doublet model~\cite{2HD}.

  The physical scalar mass can be estimated by the chiral Lagrangian
  analysis~\cite{CC:EHS,CC:HG}.  At the lowest order, 
\begin{equation}\label{pi:mass}
{m_{\pi_p}}^2 = 2c_1\sqrt{2}\frac{4\pi f}{f^\prime}v^2h,
\end{equation}
where $f^\prime$ is the scalar VEV , which is constrained together with f by
\begin{equation}\label{f:fp:v2}
f^2+{f^\prime}^2=v^2;
\end{equation}
and $h=(h_++h_-)/2$, is the average technifermion Yukawa coupling of 
$h_+$ (Yukawa coupling to $p$) and $h_-$ (Yukawa coupling to $m$).
$c_1$ is an undetermined coefficient in the chiral expansion, but of 
order unity by naive dimensional analysis~\cite{NDA}. We set the value
of it to be 1, leaving its uncertainty in that of h since they always
appear together when we work in the lowest order.

  The effective one-loop potential for the Higgs field $\sigma$, which
is the isoscalar component of $\phi$, has the following
form~\cite{CC:EHS,CC:HG},
\begin{equation}\label{potential}
V(\sigma)=\frac{1}{2}{M_\phi}^2\sigma^2 +\frac{\lambda}{8}\sigma^4 
-\frac{1}{64\pi^2}\left[3h_t^4
+N(h_+^4+h_-^4))\right]\sigma^4\log{\left(\frac{\sigma^2}{\mu^2}\right)}
-8\sqrt{2}c_1\pi f^3h\sigma,
\end{equation}
where $h_t$ is the top quark Yukawa coupling, $N=4$,
and $\mu$ is an arbitrary renormalization scale. The first three 
terms in equation (\ref{potential}) are standard one loop 
Coleman--Weinberg terms~\cite{Coleman:Weinberg}.
The last term enters the potential through the technicolor interactions. 

 Apart from the Standard Model parameters, we have four additional
parameters in this model: $(M_\phi, \lambda, h_+, h_-)$. Two limits have
been studied in the literature~\cite{EHS},\cite{CC:EHS}:
{\it (i)} the limit in which $\lambda$ is small and can be neglected; 
and {\it (ii)} the limit in which $M_\phi$ is small and can be
neglected.  The nice thing of working in these two limits is that at the
lowest order the phenomenology depends on the average of $h_+$ and $h_-$
not the difference of them. Let us look at two limits of this potential:\\

{\it (i)  $\lambda\approx 0 $, assuming the $\phi^4$ coupling is
small and can be neglected.}\\

We assume the Higgs field $\sigma$ has no VEV, and therefore terms
in the potential that are linear in $\sigma$ should vanish:

\begin{equation}
 V^\prime(\sigma)=0, 
\end{equation}
or
\begin{equation}\label{no:VEV:1}
{\widetilde{M}_\phi}^2f^\prime=8\sqrt{2}c_1\pi hf^3,
\end{equation}
where the shifted scalar mass $\widetilde{M}_\phi$ is connected to the
unshifted mass $M_\phi$ by
\begin{equation}\label{shifted:mass}
{\widetilde{M}_\phi}^2=M_\phi^2
+\left(\frac{44}{3}\right)\frac{1}{64\pi^2}\left[3h_t^4
+2Nh^4\right]{f^\prime}^2.
\end{equation}
In deriving the above two equations, we have defined the 
renormalized $(\phi^{\dag}\phi)^2$ coupling as
$\lambda_r=V^{\prime\prime\prime\prime}(f^\prime)/3$ to remove the $\mu$
dependence. For simplicity, we also set $h_+=h_-$ in
eq. (\ref{shifted:mass}).  By using the shifted scalar mass, we can
absorb radiative corrections which affect the phenomenology of the
charged scalar. However, these corrections still appear in the mass of
the $\sigma$ field, which is determined by $V^{\prime\prime}(f^\prime)$,

\begin{equation}
 { m_\sigma}^2= \widetilde{M}_\phi^2+
\left(\frac{64}{3}\right)\left(\frac{1}{64\pi^2}\right)\left[3h_t^4
+2Nh^4\right]{f^\prime}^2.
\end{equation}

In this limit, the phenomenology can be described in terms of
$(\widetilde{M}_\phi,h)$, since $h_t$ can be expressed in terms of $f$ and
$f^\prime$ ($h_t=\sqrt{2}m_t/f^\prime$). This parameterization was adopted
previously in the literature \cite{EHS}-\cite{CC:EHS:YM}.  Alternatively, we
can trade the unphysical parameter $\widetilde{M}_\phi$ for a physical
parameter, e.g., the mass of the isoscalar field, $m_\sigma$. Then the
free parameters will be two physical quantities: $(m_\sigma, h)$. \\

{\it(ii) $M_\phi\approx 0$, assuming the scalar mass is small 
and can be neglected. }\\

As in the case of limit {\it (i)}, we assume the Higgs field has no 
vacuum expectation value, in other words,
$V^\prime(\sigma)=0$ 
so that
\begin{equation}\label{no:VEV:2}
\frac{\tilde{\lambda}}{2}{f^\prime}^3=8\sqrt{2}c_1\pi hf^3,
\end{equation}
where the shifted coupling $\tilde{\lambda}$ is defined by

\begin{equation}
 \tilde{\lambda}=\lambda+\frac{11}{24\pi^2}\left[3h_t^4 + 2Nh^4\right].
\end{equation}
The same renormalization scheme as that in limit {\it  (i)} is used.
The effects of radiative corrections are absorbed 
into the shifted coupling $\tilde{\lambda}$ but still manifest in the
$\sigma$ mass, which is given by 

\begin{equation}
{m_\sigma}^2=\frac{3}{2}\tilde{\lambda}{f^\prime}^2 
    - \frac{1}{8\pi^2}\left[3h_t^4+2Nh^4\right]{f^\prime}^2.
\end{equation}
In this limit, we can choose $(\tilde{\lambda},h)$ to be our free
parameters as in refs. \cite{EHS}-\cite{CC:EHS:YM}, or again 
use $(m_\sigma,h$).\\

 We should keep in mind that these results are only valid in the part
of the parameter space where the technifermion masses ($\approx
hf^\prime$) are much smaller than the technicolor scale ($\approx 4\pi f$).
If the technifermions are heavier than this scale, the chiral
$SU(2)_L \times SU(2)_R$ will cease to be an approximate symmetry of the
theory and consequently the effective chiral lagrangian analysis will
not make sense.

%%%%%%%%%%%%%%%%%%%%%%%%%%%%%%%%%%%%%%%%%%%%%%%%%%%%%%%%%%%%%%%%%%%%%%%%%%%%%%%

\section{$B \to X_s \mu^+ \mu^-$ in the model}

  As mentioned earlier, in addition to the one-loop graphs of the 
Standard Model, additional one-loop graphs with $\pi_p$ as internal
particles are present in this model (figures
\ref{smdiag},\ref{tcsdiag}).
The scalar mass can be derived in limit {\it (i)} by using equations 
(\ref{f:fp:v2}) and (\ref{no:VEV:1}) to evaluate equation
(\ref{pi:mass}), and in limit {\it (ii)} by 
similarly combining equations (\ref{f:fp:v2}),(\ref{no:VEV:2}) and 
(\ref{pi:mass}) .

The inclusive rate for the meson level process $B \to X_s \mu^+\mu^-$
may be approximated by the rate for the free quark transition $b \to s
\mu^+ \mu^-$~\cite{Falk}, provided that the invariant mass of the
dilepton pair is not near any resonances in the charm system such as the
$\psi$. Following  refs. \cite{CDF} and \cite{Cho},  we restrict our 
analysis to the dilepton invariant mass regions
\begin{equation}\label{mregion}
m_{\mu^+\mu^-}\in(2m_\mu,2.9\mbox{GeV})\cup(3.3\mbox{GeV},
3.6\mbox{GeV})\cup(3.8\mbox{GeV},4.6\mbox{GeV}),
\end{equation}
to ensure the validity of the free quark approximation.  
In our calculation of the nonresonant $B \to X_s \mu^+ \mu^-$ 
branching fraction in the above mentioned disjoint
dilepton mass intervals, we adopt the formalism from ref.~\cite{Cho}.
For the reader' convenience, details can be found in the appendix.

\begin{figure}
\vskip -.5in
\centerline{\epsfig{file=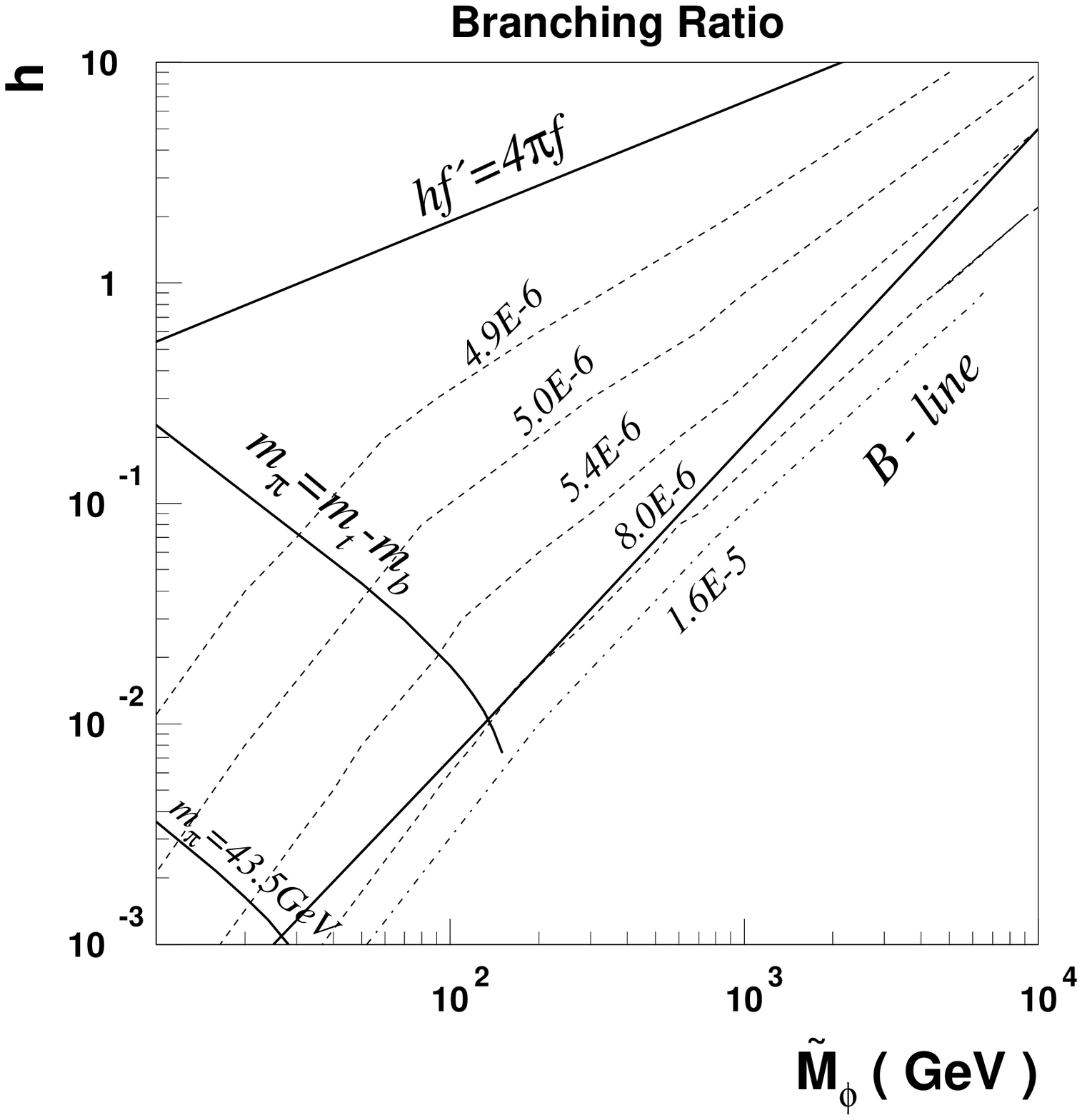,height=9cm,width=9cm}}
\vskip -1em
\caption{\label{mm1}
   Contours of $\mbox{BR}(B \to X_s \mu^+ \mu^-)_{\mbox{NR}}$
   in the  $(h,\widetilde{M}_\phi)$ plane in limit {\it (i)}.
   The allowed parameter space is bordered by B-line, $hf^\prime=4\pi f$,
   and $m_\pi=43.5$GeV.  The exclusive limit on the nonresonant 
   branching ratio from CDF,  $1.9\times 10^{-5}$ lies outside the 
   allowed region.}
\end{figure}

\begin{figure}
\vskip -.5in
\centerline{\epsfig{file=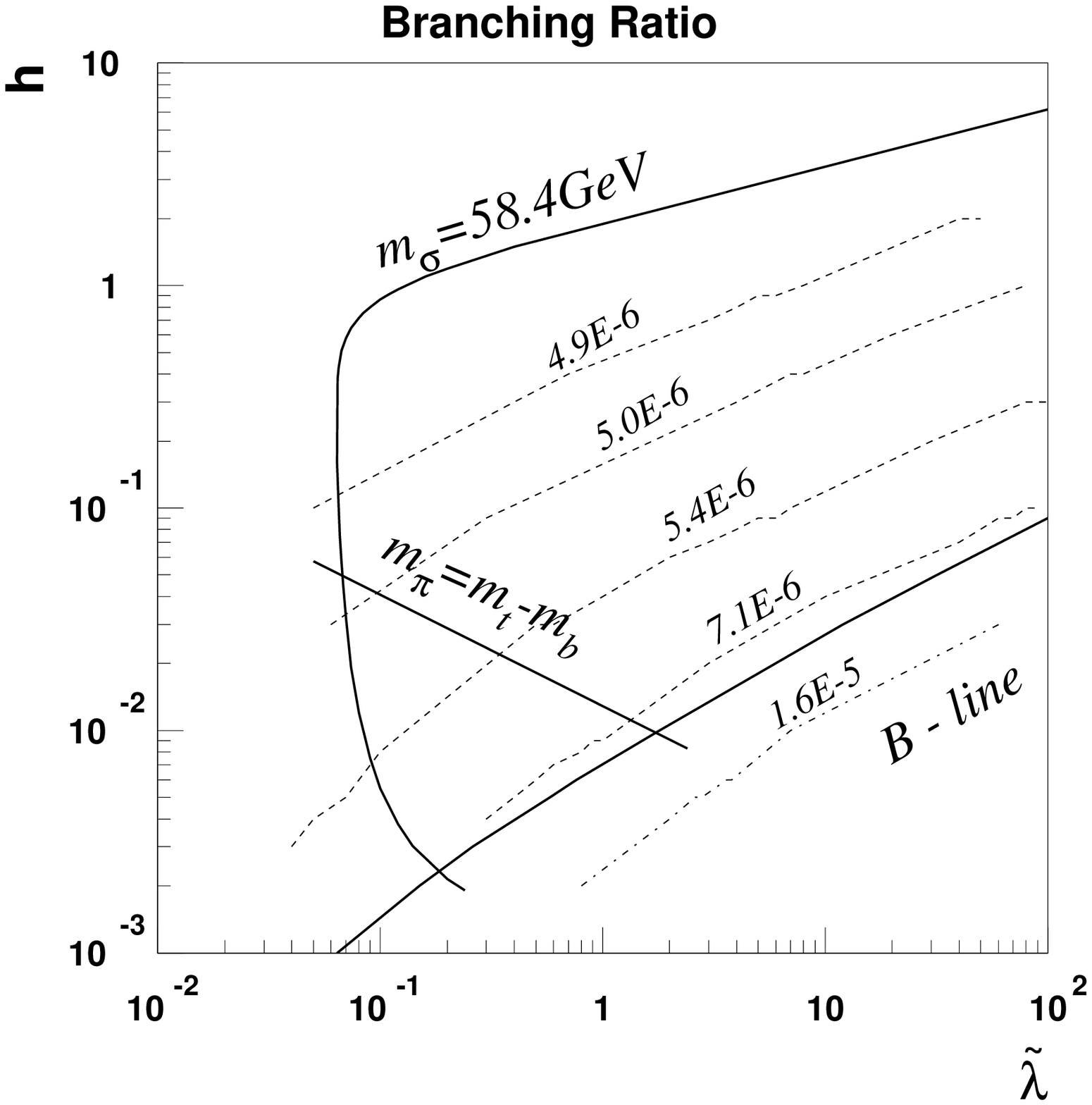,height=9cm,width=9cm}}
\vskip -1em
\caption{\label{mm2}
  Contours of $\mbox{BR}(B \to X_s \mu^+ \mu^-)_{\mbox{NR}}$ 
  in the $(h,\tilde{\lambda})$ plane in limit {\it (ii)}.
  The allowed parameter space is confined between the B-line and 
   $m_\sigma=58.4$GeV.  The exclusive limit on the nonresonant 
   branching ratio from CDF,  $1.9\times 10^{-5}$ lies outside the 
   allowed region.}
\end{figure}

\begin{figure}
\vskip -.5in
\centerline{\epsfig{file=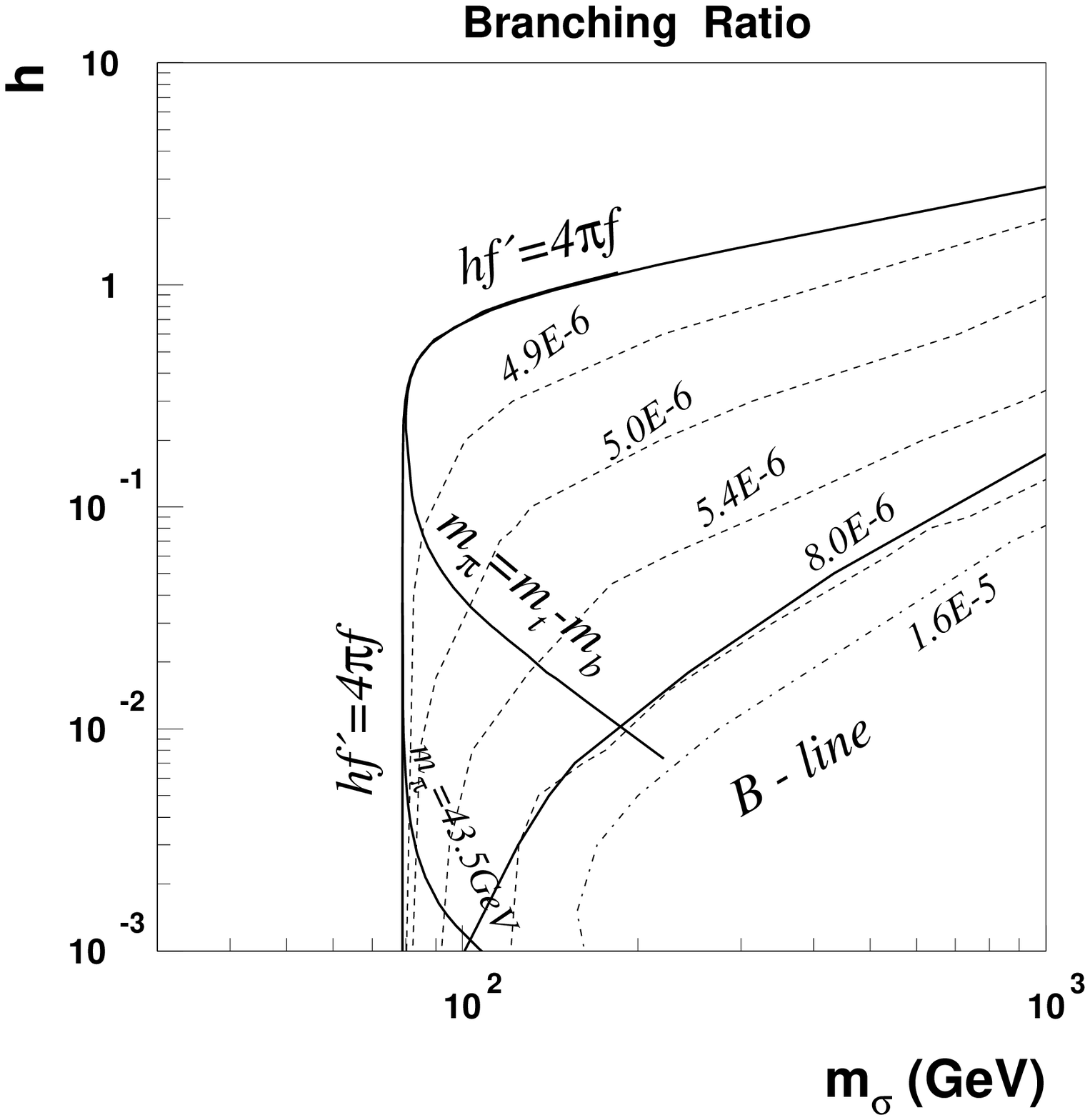,height=9cm,width=9cm}}
\vskip -1em
\caption{\label{sigma1}
  Contours of $\mbox{BR}(B \to X_s \mu^+ \mu^-)_{\mbox{NR}}$
  in the $(h, m_\sigma)$ plane in limit {\it (i)}.  The allowed parameter 
   space is bounded by B-line, $hf^\prime=4\pi f$ and $m_\pi=43.5$GeV.
  The exclusive limit on the nonresonant branching ratio from CDF, 
  $1.9\times 10^{-5}$ lies outside the allowed region.}
\end{figure}

\begin{figure}
\vskip -.5in
\centerline{\epsfig{file=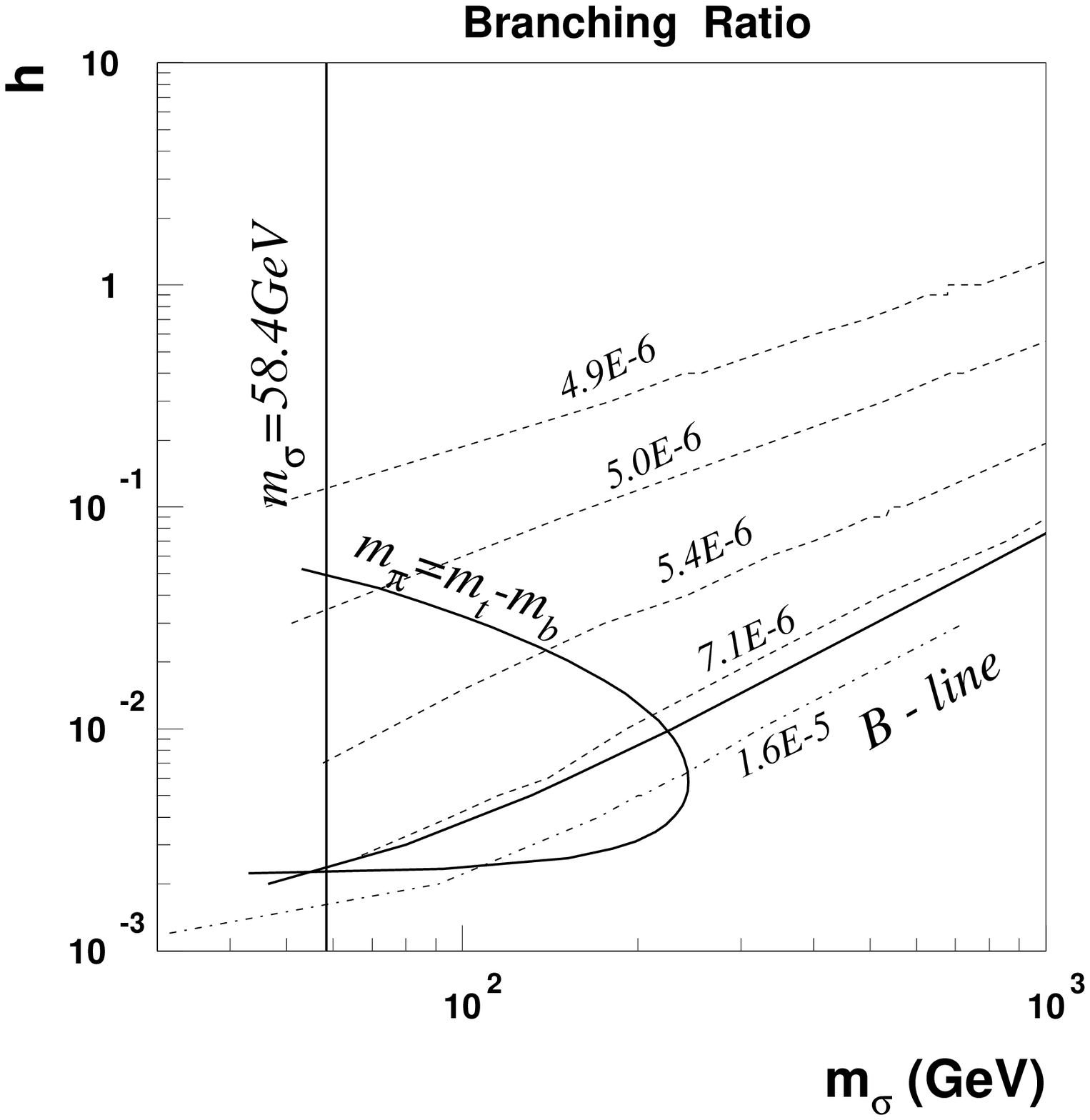,height=9cm,width=9cm}}
\vskip -1em
\caption{\label{sigma2}
   Contours of $\mbox{BR}(B \to X_s \mu^+ \mu^-)_{\mbox{NR}}$ 
   in the $(h, m_\sigma)$ plane in limit {\it (ii)}.
    The allowed parameter space is confined between B-line and 
   $m_\sigma=58.4$GeV.
   The exclusive limit on the nonresonant branching ratio from CDF, 
   $1.9\times 10^{-5}$ lies outside the allowed region.}
\end{figure}

The branching ratio of 
$b \to s \mu^+ \mu^-$ may be normalized to the semileptonic ratio,
$b \to c e\bar{\nu}$, to cancel the uncertainties arising from
the KM angles,
\begin{equation}
\mbox{BR}(b \to s \mu^+ \mu^-)
=\frac{\Gamma(b \to s \mu^+ \mu^-)}{\Gamma(b \to c e \bar{\nu})}
\mbox{BR}(b \to c e \bar {\nu}).
\end{equation}
Contours of the nonresonant branching ratio in both limits of the model
are plotted in figures \ref{mm1} to \ref{sigma2}.\\

%%%%%%%%%%%%%%%%%%%%%%%%%%%%%%%%%%%%%%%%%%%%%%%%%%%%%%%%%%%%%%%%%%%%%%%%%%%%%%%%  

%%%%%%%%%%%%%%%%%%%%%%%%%%%%%%%%%%%%%%%%%%%%%%%%%%%%%%%%%%%%%%%%%%%%%%%%%%%%%
%%%%%%%%%%%%%%%%%%%%%%%%%%%%%%%%%%%%%%%%%%%%%%%%%%%%%%%%%%%%%%%%
%%%%%%%%%%%%%%%%%%%%%%%%%%%%%%%%%%%%%%%%%%%%%%%%%%%%%%%%%%%%%%%%%%%%%%%%%
%%%%%%%%%%%%%%%%%%%%%%%%%%%%%%%%%%%%%%%%%%%%%%%%%%%%%%%%%%%%%%%%%%%%%%%%

In the {\it ``Conventional''} Parameterization:
we show the nonresonant branching ratio and the allowed parameter space
in the $(h,\widetilde{M}_\phi)$ plane in figure \ref{mm1}, and in the
$(h,\tilde{\lambda})$ plane in figure \ref{mm2}. The allowed region 
in figure \ref{mm1} is bounded by the ``B-line'', and $hf^\prime=4\pi f$
while in figure \ref{mm2} is bounded by the ``B-line'' and 
$m_\sigma=58.4$GeV. The region to the right of the ``B-line''  is 
excluded by the experimental constraints on 
$B^0 - \bar{B^0}$ mixing~\cite{EHS}.  In figure \ref{mm1},
The chiral Lagrangian analysis breaks down~\cite{CC:HG} above the 
$hf^\prime = 4 \pi f$ line . In figure \ref{mm2}, the constraint
on the mass of the isoscalar from LEP~\cite{LEP:mass} excludes the
region above and to the left of the curve 
$m_\sigma = 58.4$GeV~\cite{CC:EHS:YM}. 

In both figures, we also show the curve $m_{\pi_p}=m_t-m_b$.  Below and
to the left of this curve, $\pi_p$ is lighter than the top quark, and
the decay rate of $t\to \pi_p b$ is given by 
\begin{displaymath}
\Gamma(t\to \pi_p b)=\frac{1}{4\pi}\left[\frac{f}{f^\prime}\right]^2
                     \frac{(m_t^2-m_{\pi_p}^2)^2}{m_tv^2}.
\end{displaymath}
The branching fraction of the top decays into $W$ and $b$ measured by
CDF is $\mbox{BR}(t 
\to Wb)=0.87^{+0.13}_{-0.30}(stat.)^{+0.13}_{-0.11}(syst.)$~\cite{twb}.
The Standard Model value for $\Gamma(t\to Wb)$ is 1.6GeV.  At $2\sigma$,
the CDF data indicates that $\Gamma(t\to \pi_pb)<10.42$GeV.  This will
further constrain the parameter space.  However, because the
uncertainties in the measured $\mbox{BR}(t \to Wb)$ are too big at the
moment, we do not attempt to make any definite claim on the mass of
$\pi_p$ or to constrain parameter space further but will wait for
more accurate data.  In any case, the lower limit on
the mass of $\pi_p$ from the experimental search for charged Higgs or 
technipions~\cite{lower:pion} excludes the region below the line
$m_{\pi_p}=43.5$GeV in figure \ref{mm1}.  In figure \ref{mm2}, the
physical scalar is heavier than the experimental lower bound in the
entire allowed space.

In terms of the {\it physical} parameterization, $(m_\sigma, h)$, in
both  limits
of the model, we plot contours of the nonresonant branching ratio in
figures \ref{sigma1} and \ref{sigma2}. In this case, the boundaries of
the allowed parameter space are the same as in the  
``conventional'' parameterization.  There are some
advantages of this choice of free parameters. First, it enables us to 
visualize what the parameter space looks like for a fixed not-so-small
$M_\phi$ (or $\lambda$); second, when the limit 
on the isoscalar mass changes, we simply need to move the
vertical line $m_\sigma=58.4\mbox{GeV}$ to get the updated parameter
space without carrying out a lengthy computation.
 
In figures \ref{mm1} and \ref{sigma2}, as we move from upper left
to lower right in the allowed region of the parameter space , we find 
that the nonresonant branching ratio 
increases from the Standard Model value, $4.9 \times 10^{-6}$, to about 
$8.0 \times 10^{-6}$. This corresponds to a maximum enhancement relative
to the value in the Standard Model about $60\%$. The exclusive limit on
the nonresonant branching fraction from CDF~\cite{CDF},
$1.9 \times 10^{-5}$, lies below the region allowed by the
$B^0 - \bar{B^0}$ mixing. 
The parameter space is not constrained further by $B \to X_s \mu^+\mu^-$.  
It is also not surprising to see that the ``B-Line'' and the contours look similar, 
since the $B^0 - \bar{B^0}$ mixing and $b \to s \mu^+\mu^-$ involve the same 
$\pi_p - t$ loop.

The branching ratio depends on the sign and  magnitude of the Wilson
coefficient $C_7$ of the electromagnetic operator in the effective 
Hamiltonian for the $B \to X_s \ell^+ \ell^-$ decay~\cite{Grinstein}.  
The uncertainty in the calculation of 
$C_7$ (about $15\%$)~\cite{Grinstein} can shift
the lines $\mbox{BR}(B \to X_s\mu^+\mu^-)_{\mbox{\scriptsize NR}}=1.6 
\times10^{-5}$ by at most $5\%$ only, which will not move this line 
above the ``B-line''. The corresponding shift in $\mbox{BR}(B \to
X_s\mu^+\mu^-)_{\mbox{\scriptsize NR}}=8.0 \times10^{-6}$ is about $1\%$. 

  We also compute the process $B \to X_s e^+ e^-$ in the
model.  The maximal enhancement of the nonresonant branching ratio
relative to its SM counterpart is about $20\%$,  which is comparable to the
10-20$\%$ uncertainties in the SM calculation.  Experimentally,  it is hard
to distinguish this model from the SM in the $B \to X_s e^+ e^-$ decay 
channel.

  However, given the fact that CLEO is upgrading their detector and the
 sensitivity of CDF Run II,  the $60\%$ enhancement of the 
$B \to X_s \mu^+ \mu^-$ nonresonant branching ratio in technicolor 
with scalars will enable them to distinguish the SM from this model in
 the $B \to X_s \mu^+ \mu^-$ channel.

%%%%%%%%%%%%%%%%%%%%%%%%%%%%%%%%%%%%%%%%%%%%%%%%%%%%%%%%%%%%%%%%%%%%%%%%%%%%%%
\section[short title]{Conclusions}

To extend the phenomenology of the technicolor with scalars model, we have
computed the inclusive decay $B \to X_s \mu^+ \mu^- $ in it.  The model 
predicts a significant
enhancement of the branching ratio in part of its parameter space. 
While current experiments can not quite see it, CDF Run II and the 
upgraded CLEO detector would be sensitive enough to detect it, if nature 
does not trifle with us as to make the branching ratio much smaller than
the prediction of the Standard Model; on the other hand, if some completely
different physics makes the branching ratio too small, the experiments will
still set a new upper limit.  Then, we would be able to determine the 
model's viability.

%%%%%%%%%%%%%%%%%%%%%%%%%%%%%%%%%%%%%%%%%%%%%%%%%%%%%%%%%%%%%%%%%%%%%%%%%%%%%%%%

\section*{Acknowledgments}

The author would like to thank E. H. Simmons for helpful 
discussions and comments the manuscript, and D. Loomba and B. Zhou
for their help with diagrams. {\em This work was supported in part by the
Department of Energy under grant DE-FG02-91ER40676.}

%%%%%%%%%%%%%%%%%%%%%%%%%%%%%%%%%%%%%%%%%%%%%%%%%%%%%%%%%%%%%%%%%%%%%%%%%%%%%
\renewcommand{\theequation}{A\arabic{equation}}
\setcounter{equation}{0}
\section*{Appendix}

This appendix contains details on the calculation of the $B \to X_s 
\ell^+\ell^- (\ell = e,\mu)$ branching ratio. We adopt the formalism 
from ref~\cite{Cho}.
 For the reader's convenience, we give the explicit formulas below.

Define the rescaled lepton energies in the b quark rest frame,
\begin{equation}
y_+=\frac{2E_{\ell^+}}{m_b}\ \ \mbox{and}\ \ \ y_-=\frac{2E_{\ell^-}}{m_b}
\end{equation}
and the rescaled invariant dilepton mass $\hat{s}=y_+-y_--1$.  The
differential decay rate is,
\begin{eqnarray}
\frac{d^2\Gamma(b\to s\ell^+\ell^-)}{dy_+dy_-}&=&
               \frac{G_F^2m_b^5|K_{ts}^*K_{tb}|^2}{16\pi^3}
               \left(\frac{\alpha_{\scriptstyle EM}}{4\pi}
                \right)^2\Bigg\{\bigg[y_+(1-y+)
                   \nonumber\\
 & &        +y_-(1-y_-)\bigg]
               (|C^{\mbox{\scriptsize eff}}_9(\hat{s})|^2
               +C_{10}^2)\nonumber \\
 & &        +\frac{4}{\hat{s}}\left[\hat{s}(1-\hat{s})+(1-y_+)^2+(1-y_-)^2
              +\frac{2m_{\ell}^2}{\hat{s}m_b^2}\right](C^{\mbox{{\scriptsize
                eff}}}_7)^2\nonumber\\
 & &        +4(1-\hat{s})C_7^{\mbox{{\scriptsize 
          eff}}}\mbox{Re}(C_9^{\mbox{{\scriptsize eff}}}(\hat{s}))\nonumber\\
 & &        +2(y_+-y_-)C_{10}\left[2C_7^{\mbox{{\scriptsize
                eff}}}+\hat{s}\mbox{Re}(C_9^{\mbox{{\scriptsize
                eff}}}(\hat{s}))\right]\Bigg\},
\end{eqnarray}
where,
\begin{eqnarray}
C_7^{\mbox{{\scriptsize eff}}}&=&
               C_7(m_{\scriptstyle W})\eta^{16/23}
              +\frac{8}{3}C_8(m_{\scriptstyle W})(\eta^{14/23}-\eta^{16/23})
              +\sum_{i=1}^8h_i\eta^{a_i},\label{C7eff}\\
C_9^{\scriptsize eff}&=&\left(\frac{\pi}{\alpha_s(m_{\scriptstyle W})}
              +\frac{\omega(\hat{s})}{\eta}\right)(-0.1875
              +\sum_{i=1}^8p_i\eta^{a_i+1}) \nonumber\\
  & &        +\frac{Y(x_t)+Y^{\scriptsize TCS}}{\sin^2{\theta}}
              -4(Z(x_t)+Z^{\scriptsize TCS})+(E(x_t)
              +E^{\scriptsize TCS})(0.1405+\sum_{i=1}^8q_i\eta^{a_i+1})
                \nonumber\\
  & &        +1.2468+\sum_{i=1}^8\eta^{a_i}\left[r_i
              +s_i\eta+t_ih(\frac{m_c}{m_b},\hat{s})
              +u_ih(1,\hat{s})+v_ih(0,\hat{s})\right].\label{C9eff}\\
C_{10}&=&-\frac{Y(x_t)+Y^{\mbox{\scriptsize TCS}}}{\sin^2{\theta}}.\label{c10}
\end{eqnarray}
In the above, $\eta$ is defined by 
$\eta=\alpha_s(m_{\scriptstyle W})/\alpha_s(m_b)$, and
the dimension-8 vectors are given by
\begin{eqnarray}
a_i&=&(0.6087, 0.6957,0.2609,-0.5217,0.4086,-0.4230,-0.8994,0.1456),
       \nonumber\\
h_i&=&(2.2996,-1.0880,-0.4286,-0.0714,-0.6494,-0.0380,-0.0186,-0.0057,
       \nonumber\\
p_i&=&(0,0,-0.3941,0.2424,0.0433,0.1384,0.1648,-0.0073),
       \nonumber\\
q_i&=&(0,0,0,0,0.0318,0.0918,-0.2700,0.0059),
       \nonumber\\
r_i&=&(0,0,0.8331,-0.1219,-0.1642,0.0793,-0.0451,-0.1638),\\
s_i&=&(0,0,-0.2009,-0.3579,0.0490,-0.3616,-0.3554,0.0072),
       \nonumber\\
t_i&=&(0,0,1.7143,-0.6667,0.1658,-0.2407,-0.0717,0.0990),
       \nonumber\\
u_i&=&(0,0,0.2857,0,-0.2559,0.0083,0.0180,-0.0562),
       \nonumber\\
v_i&=&(0,0,0.1429,0.1667,-0.1731,-0.1120,-0.0178,-0.0067).
       \nonumber
\end{eqnarray}
And the functions $h(z,\hat{s})$, and $\omega(\hat{s})$ are given by
\begin{eqnarray}
h(z,\hat{s})&=&-\frac{8}{9}\log{z}+\frac{8}{27}+\frac{4}{9}x\nonumber\\
 & &       -\frac{2}{9}(2+x)\sqrt{|1-x|}\left\{\begin{array}{l}
                  \log\left|\frac{\sqrt{1-x}+1}{\sqrt{1-x}-1}\right|-i\pi,\ 
                  \ \ \ \ \mbox{for}\  x\equiv 4z^2/\hat{s}<1,\nonumber\\
                 2\arctan{(1/\sqrt{x-1})},\ \mbox{for}\ x\equiv
                4z^2/\hat{s}>1; \end{array} \right.\nonumber\\
\omega(\hat{s})&=&-\frac{4}{3}\mbox{Li}_2(\hat{s})
        -\frac{2}{3}\log{(\hat{s})}\log{(1-\hat{s})}-\frac{2}{9}\pi^2
        -\frac{5+4\hat{s}}{3(1+2\hat{s})}\log{(1-\hat{s})}\nonumber\\
& &     -\frac{2\hat{s}(1+\hat{s})(1-2\hat{s})}{3(1-\hat{s})^2(1
            +2\hat{s})}\log{(\hat{s})}
        +\frac{5+9\hat{s}-6\hat{s}^2}{6(1-\hat{s})(1+2\hat{s})}.
\end{eqnarray}
The differential rate is integrated over the dilepton invariant
mass region given by (\ref{mregion}) to get the partial with for 
$b \to s \ell^+ \ell^-$, which then is normalized to the semileptonic
rate
\begin{equation}
\Gamma(b \to
ce^+\nu)=\frac{G_F^2m_b^5|K_{cb}|^2}{192\pi^3}g(\frac{m_c}{m_b})
\left\{1-\frac{2\alpha_s(m_b)}{3\pi}\left[(\pi^2-\frac{31}{4})
(1-\frac{m_c}{m_b})^2+\frac{3}{2}\right]\right\}
\end{equation}
to cancel the uncertainties in the KM angles;
here $g(z)=1-8z^2+8z^6-z^8-24z^4\log{z}$.
Adopting the values: $m_c=1.3$GeV,
$m_b=4.7$GeV, $m_t=176$GeV,
$\alpha_s(m_Z)=0.118$ and $|K^*_{ts}K_{tb}/K_{cb}|^2=0.95$,
the nonresonant branching ratios are given by
\begin{equation}\label{bree}
\begin{array}{l}
\mbox{BR}(B \to X_s e^+ e^-)_{\mbox{\scriptsize NR}}=
                       3.0 \times 10^{-7}[5.5 
                       + 2.3R_7^2 +17.6R_Y^2\\
\ \ \ \ \ \ \ \ \ \ \  +3.7 R_Z^2 -2.1R_7R_Y +1.4R_7R_Z -11.5R_YR_Z\\
\ \ \ \ \ \ \ \ \ \ \  +4.6R_7 +8.1R_Y -5.3R_Z],
\end{array}
\end{equation}
\begin{equation}\label{brmm}
\begin{array}{l}
\mbox{BR}(B \to X_s \mu^+ \mu^-)_{\mbox{\scriptsize NR}}=
                      3.0 \times 10^{-7}[2.9 
                      + 0.8R_7^2 +17.5R_Y^2\\
\ \ \ \ \ \ \ \ \ \ \  +3.7 R_Z^2 -2.1R_7R_Y +1.4R_7R_Z -11.4R_YR_Z\\
\ \ \ \ \ \ \ \ \ \ \  +0.7R_7 +8.1R_Y -5.3R_Z],
\end{array}
\end{equation}
where,
\begin{eqnarray}
R_7&=&\frac{C_7(m_{\scriptstyle W})_{\mbox{{\scriptsize SM}}}
 +C_7(m_{\scriptstyle W})_{\mbox{{\scriptsize 
 TCS}}}}{C_7(m_{\scriptstyle W})_{\mbox{{\scriptsize SM}}}}
,\nonumber\\
R_Y&=&\frac{Y(x_t)+Y^{\mbox{{\scriptsize TCS}}}}{Y(x_t)},\\
R_Z&=&\frac{Z(x_t)+Z^{\mbox{{\scriptsize TCS}}}}{Z(x_t)}.\nonumber
\end{eqnarray}
Here, TCS stands for the extra contributions from technicolor with
scalars, and SM stands for the Standard Model.  $R_7=R_Y=R_Z=1$ gives
the Standard Model nonresonant branching ratio $7.3\times 10^{-6}$ for 
$B \to X_s e^+ e^-$ and $4.9\times 10^{-6}$ for $B \to X_s\mu^+\mu^-$.
The extra contributions in TCS are
\begin{eqnarray}
Y^{\mbox{{\scriptsize TCS}}}&=
&-\frac{1}{8}\left(\frac{f}{f^\prime}\right)^2
                 x_tf_5\left(\frac{m_t^2}{m_{\pi_p^\pm}^2}\right),
\nonumber\\
Z^{\mbox{{\scriptsize TCS}}}&=
&-\frac{1}{8}\left(\frac{f}{f^\prime}\right)^2
              x_t f_5\left(\frac{m_t^2}{m_{\pi_p^\pm}^2}\right)
-\frac{1}{72}\left(\frac{f}{f^\prime}\right)^2
              f_6\left(\frac{m_t^2}{m_{\pi_p^\pm}^2}\right),
\label{Ztcs}
\end{eqnarray}
where $x_t$ is defined by $x_t=(m_t/m_{\scriptstyle W})^2$.
  
The functions $Y(x)$, $Z(x)$, and $E(x)$ appearing in $R_Y$, $R_Z$,
$C_9^{\scriptsize eff}$, and $C_{10}$ are determined by the following:

\begin{eqnarray}
Y(x)&=&\frac{4x-x^2}{8(1-x)}+\frac{3x^2}{8(1-x)^2}\log{x},\nonumber\\
Z(x)&=&\frac{108x-259x^2+163x^3-18x^4}{144(1-x)^3}\nonumber\\
 &\ &+\frac{-8+50x-63x^2-6x^3+24x^4}{72(1-x)^4}\log{x},\\
E(x)&=&\frac{18x-11x^2-x^3}{12(1-x)^3}
       -\frac{4-16x+9x^2}{6(1-x)^4}\log{x}.\nonumber
\end{eqnarray}

$C_7(m_{\scriptstyle W})$ is the Wilson coefficient evaluated at the
scale $m_{\scriptstyle W}$,

\begin{eqnarray}
C_7(m_{\scriptstyle W})_{\mbox{{\scriptsize 
     SM}}}&=&\frac{1}{2}(-\frac{x}{2}f_1(x))\label{C7sm}\nonumber\\
C_7(m_{\scriptstyle W})_{\mbox{{\scriptsize TCS}}}&=
&\frac{1}{6}\left(\frac{f}{f^\prime}\right)^2
                \left[-f_2\left(\frac{m_t^2}{m_{\pi_p^\pm}^2}\right)
          +\frac{1}{2}\frac{m_t^2}{m_{\pi_p^\pm}^2}
          f_1\left(\frac{m_t^2}{m_{\pi_p^\pm}^2}\right)\right].\label{C7tcs}
\end{eqnarray}
Here, $C_7(m_{\scriptstyle W})_{\mbox{{\scriptsize TCS}}}$ has the same
form as in a type-I two Higgs doublet model~\cite{Grinstein}.
 
Finally, the f functions appearing in (\ref{Ztcs})
and (\ref{C7tcs}) are expressed as,

\begin{eqnarray}
f_1(x)&=&\frac{-7+5x+8x^2}{6(1-x)^3}-\frac{2x-3x^2}{(1-x)^4}\log{x}\nonumber\\
f_2(x)&=&\frac{3x-5x^2}{2(1-x)^2}+\frac{2x-3x^2}{(1-x)^3}\log{x}\nonumber\\
f_5(x)&=&\frac{x}{1-x}+\frac{x}{(1-x)^2}\log{x}\nonumber\\
f_6(x)&=&\frac{38x-79x^2+47x^3}{6(1-x)^3}+\frac{4x-6x^2+3x^4}{(1-x)^4}\log{x}.
\end{eqnarray}
%%%%%%%%%%%%%%%%%%%%%%%%%%%%%%%%%%%%%%%%%%%%%%%%%%%%%%%%%%%%%%%%%%%%%%%%%%%%%

%%%%%%%%%%%%%%%%%%%%%%%%%%%%%%%%%%%%%%%%%%%%%%%%%%%%%%%%%%%%%%%%%%%%%%%%%%

\end{document}